\documentclass[prb,showpacs,twocolumn]{revtex4}
\usepackage{amsmath}
\usepackage{amssymb}
\usepackage{bm}
\usepackage{epsfig}
\usepackage{graphicx}
\usepackage{color}
\usepackage{natbib}
\usepackage{hyperref}

\renewcommand{\Re}{\mathop{\rm Re}\nolimits}
\renewcommand{\Im}{\mathop{\rm Im}\nolimits}

\renewcommand{\i}{{\mathrm i}}
\newcommand{\e}{{\mathrm e}}
\begin{document}
\title{
Non-linear emission spectra of quantum dots strongly coupled to photonic mode}

\author{A.N. Poddubny, M.M. Glazov, N.S. Averkiev}

\affiliation{Ioffe Physical-Technical Institute RAS, 26 Polytekhnicheskaya, 194021 St.-Petersburg, Russia}
\begin{abstract}
A theory of optical emission of quantum dot arrays in quantum microcavities is developed. The regime of the strong coupling between the quantum dots and photonic mode of the cavity is considered. The quantum dots are modeled as two-level systems. In the low pumping (linear) regime the emission spectra are mainly determined  by the superradiant mode where the effective dipoles of the dots oscillate in phase. In the non-linear regime the superradiant mode is destroyed and the emission spectra are sensitive to the parity of quantum dot number. Further increase of the pumping results in the line width narrowing being an evidence of the lasing regime.
\end{abstract}

\date{\today}

\maketitle

\section{Introduction}\label{sec:intro}

The light-matter coupling on the nanoscale is in focus of the research during the last decade owing to the possibilities to realize quantum electrodynamical effects in the solid state and to the perspectives of the efficient emission control needed for device applications. Semiconductor microcavities with embedded quantum dots are among the most promising objects in these respects. Two regimes of the light-matter interaction in such systems can be identified. The first one, weak coupling regime, is characterized by the relatively small interaction constant as compared with the decay rates of the photonic and excitonic states. In this case, the microcavity merely modifies the quantum dot emission rate owing to Purcell effect. The second, strong coupling regime corresponds to the large coupling constant as compared with the decay rates which makes possible  the coherent energy transfer between the quantum dot and a cavity resulting in formation of the mixed, half-light -- half-matter modes called exciton-polaritons. Such a strong coupling regime was observed recently in semiconductor microcavities with quantum dots.\cite{Reithmaier2004,Khitrova2004} 

Although, in most of the experimental and theoretical studies the simplest possible situation of one quantum dot being close in energy to the cavity resonance was considered so far,
\cite{Peter2005,Henessy2007}
it is feasible to achieve the strong coupling for multiple quantum dots. \cite{kulakovskii2006} 
 Theoretical analysis \cite{JETP2009} demonstrated that, at the weak pumping where both cavity mode and quantum dots can be treated as classical harmonic oscillators the optical properties are dominated by the superradiant mode which can be considered as a collective in-phase oscillation of quantum dot dipole moments.\cite{Dicke1954,khitrova2007Nat} In perfect systems this mode  is the only one which interacts with a photon and their coupling constant is strongly enhanced as compared with the single dot case. The superradiant mode is relatively stable to the impact of disorder (i.e. spread of quantum dot resonant energies) and can be even stabilized by the allowance for the interdot tunneling.\cite{JETP2009} 

The increase of the pumping rate yields the increase of the electric field in a cavity and of the exciton dipole moment oscillation amplitudes.  Thereby, the model of the classical oscillators becomes violated and the unharmonicity effects have to be taken into account. Important questions in this respect are (i) how the emission spectra of the strongly-coupled system of quantum dots and cavity change as a function of pumping and (ii) whether the superradiant mode survives to some extent in the non-linear regime? The present paper is devoted to the theoretical study of these questions.

We develop here a theory of non-linear emission of quantum dot ensembles embedded in microcavities in the strong coupling regime. We assume that the quantum dots are relatively small so that the electron and hole forming a zero-dimensional exciton are quantized independently and no more than one electron-hole pair can occupy the dot. The incoherent pumping of the excitonic states is considered. We also take into account the effects of the quantum dot resonant energies spread. 

The main results of our paper can be summarized as follows:
\begin{enumerate}
\item The emission spectra of the single quantum dot detuned from the cavity resonance strongly depends on the pumping rate: it transforms from the doublet with dominant dot emission line to the single peak spectrum positioned near the cavity resonance with an increase of the pumping power.
\item In the case where the spread  of resonance energies of quantum dots is negligible, the non-linear emission spectra at moderate pumping rates exhibit either a single or a double peak structure depending whether the number of dots in a cavity is even or odd, respectively.

\item At relatively high pumping rates the emission shows a single line whose width decreases with an increase of the pumping being a signature of the lasing regime.
\end{enumerate}

The paper is organized as follows: Section~\ref{sec:model} describes the model assumptions and outlines the calculation procedure. We demonstrate, that in strong coupling regime with well defined polariton states, the optical spectra can be found from the kinetic equation. An advantage of this approach over the full density-matrix calculations,\cite{laussy2009} besides lower computational costs, is the possibility to derive transparent analytic answers for the optical spectra. In Sec.~\ref{sec:1QD} the developed method is applied to the case of a single quantum dot where the results known in the literature are reproduced in a simple fashion in the case of a dot being degenerate with the cavity mode. We also consider in this Section the situation of a dot detuned from the cavity mode and demonstrate the specific properties of the emission spectra. Section~\ref{sec:mes} is devoted to the case of several quantum dots in a cavity and the quantum dot number parity effects are studied there in detail. To a certain extent, our results are valid in the case of a single quantum dot with multiple size-quantization levels in the vicinity of the photonic mode resonance, studied in Ref.~\onlinecite{PhysRevB.80.125302}. Concluding remarks are presented in Sec.~\ref{sec:concl}. 
Derivation of analytical results for the microcavity with single dot is presented in Appendix~\ref{sec:A1}. The effects of exciton-exciton interactions are briefly discussed in Appendix~\ref{sec:A2}.
 
\section{Model}\label{sec:model}

We are interested in the microcavity with $N$ embedded quantum dots. It is assumed that only one photonic mode of the cavity is of importance, i.e. it is close enough to the quantum dot transition energies. Each quantum dot is described by a two-level system where the ground state corresponds to the empty quantum dot and the excited state corresponds to the dot occupied by a single electron-hole pair (exciton). Such an assumption is readily realized for the small quantum dots whose effective size is smaller as compared with the exciton Bohr radius.\cite{Laussy2006} For the purposes of the present paper the spin degrees of freedom of photons and excitons are disregarded.

Under above assumptions the Hamiltonian of the studied system can be written in the following form \cite{kavbamalas}
\begin{equation}
\label{eq:Hamiltonian}
\mathcal H = \hbar \omega_{\rm C} c^\dag c + \sum_{i=1}^N \hbar\omega_{{\rm X},i} b^\dag_i b_i^{\vphantom{\dag}} + \hbar g\sum_i (c^\dag b_i^{\vphantom{\dag}}+ cb_i^\dag).
\end{equation}
Here $\omega_{\rm C}$ is the resonance frequency of the cavity, $\omega_{{\rm X},i}$ ($i=1,\ldots, N$) are quantum dot resonance frequencies, $c^\dag$, $c$ are the creation and annihilation operators for the photon mode, respectively,  and $b^\dag_i$, $b_i^{\vphantom{\dag}}$ are the analogous operators for the quantum dot modes. Cavity mode is bosonic while quantum dots are treated as two-level systems. Therefore operators $b^\dag_i$, $b_j$ obey Fermi commutation rules for $i=j$ ($b_i^\dag b_i^{\vphantom{\dag}}+b_i^{\vphantom{\dag}}b_i^\dag=1$), while for different $i\ne j$ quantum dot operators simply commute.~\footnote{It can be easily justified assuming that the exciton creation operator is a product of Fermionic operators for electrons and holes.}   The last term in Eq.~\eqref{eq:Hamiltonian} describes the coupling between the excitons and photon, $\hbar g$ is the coupling strength. It is assumed here that the coupling constant is identical for all dots, the developed theory can be easily extended to allow for different coupling strengths.

The different nature of photon and exciton states makes it impossible to consider the dynamics of the coupled quantum dots/microcavity system in a model of classical harmonic oscillators. Indeed, the quantum dot cannot accommodate more than one exciton, although the number of photons in the cavity is not limited. At small pumping densities the excitation is transferred back and forth between the exciton and photon modes. An increase of the pumping power can result in the appearance of extra photons in the cavity. These extra photons  cannot be absorbed by the quantum dot but modify the light-matter coupling instead.

In our model we neglect completely the interaction between excitons. Clearly, since we model each dot as a two level system the number of excitons in each quantum dot is no more than one and there is no interaction within the same dot. The Coulomb interaction of excitons in different dots is neglected as well assuming that it is much smaller than the coupling constant $\hbar g$. The effects of interactions are briefly discussed in Appendix~\ref{sec:A2}.

The Hamiltonian~\eqref{eq:Hamiltonian} describes the eigenstates and the energy levels of the quantum dots in the microcavity in the absence of pumping and the non-radiative decay of the excitons and photons. We introduce the pumping rate to the $i$th exciton state, $W_i$, that is the number of excitons generated in a quantum dot per unit of time and the decay rates of the exciton and photon populations, $\Gamma_{{\rm X},i}$ ($i=1,\ldots, N$), $\Gamma_{\rm C}$, respectively. The cavity decay rate $\Gamma_{\rm C}$ defines the number of photons leaving the cavity per unit of time due to the non-zero transparency of the mirrors and the exciton population decay rates $\Gamma_{{\rm X},i}$ characterize the non-radiative processes in quantum dots. 
In the present paper we focus on the strong coupling regime where the eigenstates of the Hamiltonian~\eqref{eq:Hamiltonian} are well defined. Therefore, we assume that $\Gamma_{{\rm X},i}, \Gamma_{\rm C} \ll g$ ($i=1,\ldots, N$). The requirement of the well defined states also imposes a certain restriction on the pumping rate $W{<g^2/\Gamma_{\rm C},g^2/\Gamma_{{\rm X},i}}$, see below. 

The Hamiltonian~\eqref{eq:Hamiltonian} describes the energy transfer between the quantum dots and the microcavity and conserves the total number of particles in the system (excitons and photons). Therefore, its eigenstates can be labeled by the total number of particles, $m$, (and, of course, by other quantum numbers which take into account the degeneracy of non-interacting $m$ particle states, e.g. $m$ photons, $0$ excitons or $m-1$ photons, $1$ exciton, etc.) For instance, for a single quantum dot in the microcavity tuned exactly to the photon mode, $\omega_{\rm X}=\omega_{\rm C}$ the states are combinations 
\begin{equation}
\label{eq:jc:states}
(|m,0\rangle \pm |m-1,1\rangle)/\sqrt{2},
\end{equation}
 where $|n_{\rm C}, n_{\rm X}\rangle$ denotes the state with the definite number of photons, $n_{\rm C}$, and excitons, $n_{\rm X}$, with energies
 \begin{equation}
\label{eq:jc:energ}
 E_{m,\pm} = m \hbar \omega_{\rm C} \pm \sqrt{m} \hbar g.
 \end{equation} 
These states with different $m$ form Jaynes-Cummings ladder.\cite{jaynes1963,Shore1993} The extension of Eqs.~\eqref{eq:jc:states}, \eqref{eq:jc:energ} for the important cases of a detuned quantum dot and multiple quantum dots in a cavity are presented below in Secs.~\ref{sec:1QD} and \ref{sec:mes}, respectively. It is seen from Eq.~\eqref{eq:jc:energ} that the energy spectrum of the interacting quantum dot and microcavity depends strongly on the number of particles. The light-matter coupling strength increases with $m$.

The manifolds, i.e. the sets of states with fixed $m$,
are intermixed by the dissipative processes.
 Namely, the emission of photon results in the transition from the manifold with $m$ particles to the manifold with $m-1$ particle, while the generation of an exciton in a quantum dot leads to the increase of the number of particles by $1$, i.e. to the transition from the manifold $m$ to the manifold $m+1$. In the strong coupling regime one can describe the populations of the states in the framework of the distribution function $f_{m,\mu}$, where the subscript $\mu$ enumerates different states in the manifold [e.g., $\mu=\pm$ in the case of single dot in a cavity, see Eqs.~\eqref{eq:jc:states}, \eqref{eq:jc:energ}]. The distribution function is governed by the phenomenological  kinetic equation which has the following form
\begin{equation}
\label{eq:kinetic}
\frac{\mathrm d f_{m,\mu}}{\mathrm dt}= -D_{m,\mu} f_{m,\mu} +\\ \sum_{m'=m\pm 1,\mu'} W_{m',\mu' \to m,\mu} f_{m',\mu'},
\end{equation}
where $D_{m,\mu}$ is the total decay rate of the state $m, \mu$ caused by the transitions to lower and upper manifolds resulting from the photon escape from the cavity, non-radiative exciton decay and pumping,
\[
D_{m,\mu} = \sum_{m'=m\pm 1,\mu'} W_{m,\mu \to m\pm 1,\mu'},
\]
and $W_{m',\mu' \to m,\mu}$ are the corresponding transition probabilities. 
They can be separated into two parts, corresponding to the photon escape from the cavity, $W^{\rm C}$, and exciton generation in $i$th dot, $W^{\mathrm X,i}$:
\begin{gather}\label{eq:rates}
 W_{m,\mu \to m',\mu'}=W^{\rm C}_{m,\mu \to m',\mu'}+
\sum\limits_iW^{{\rm X},i}_{m,\mu \to m',\mu'},
\\\nonumber
W^{\rm C}_{m,\mu \to m',\mu'}=\Gamma_{\rm C}\delta_{m',m-1}
|c_{m-1,\mu';m,\mu}|^2,\\\nonumber
W^{{\rm X},i}_{m,\mu \to m',\mu'}=(\Gamma_{{\rm X},i}\delta_{m',m-1}+
W_i\delta_{m',m+1})|b^{(i)}_{m',\mu';m,\mu}|^2\:.
 \end{gather}
Here  $c_{m',\mu';m,\mu}$ and $b_{m',\mu';m,\mu}^{(i)}$ are the matrix elements of
annihilation operators of photon and exciton in $i$-th dot, respectively,
taken between the states $m,\mu$ and $m',\mu'$. It is worth to stress, that both pumping and decay events change the number of particles by one, thus coupling only neighboring manifolds: $m$, $m\pm 1$. Note, that the manifold number $m$ in Eq.~\eqref{eq:kinetic} varies from $0$ to infinity, i.e. including the ground state of the system being the state with no particles. The distribution function obeys the normalization condition
\begin{equation}
\label{eq:norm}
\sum_{m=0}^\infty\sum_\mu f_{m,\mu}=1.
\end{equation}

In the kinetic approach the emission spectra can be calculated by the Fermi golden rule
\begin{equation}
\label{eq:Fermi}
I(\omega) \propto \sum_{m,\mu,\mu'}f_{m,\mu}  W^{\rm C}_{m,\mu\to m-1,\mu'}\delta(E_{m,\mu} - E_{m-1,\mu'} -\hbar \omega ),
\end{equation}
and are determined by the rate of the photon escape from the cavity 
$W^{\rm C}_{m,\mu\to m-1,\mu'}$. Here $E_{m,\mu}$ is the energy of the corresponding state, and the $\delta$-function ensures the energy conservation. The common factor in Eq.~\eqref{eq:Fermi} containing the mirror transmission coefficients, etc., is ignored. To derive  Eq.~\eqref{eq:Fermi} we considered the process of the photon escape from the cavity 
and represented the emission intensity as a weighted with the stationary distribution function $f_{m,\mu}$ sum over all manifolds. Within the applicability of the kinetic equation the line broadening is to be small as compared with the transition energy, the condition surely satisfied in the strong coupling regime. It is seen that Eq.~\eqref{eq:Fermi} gives infinitely narrow emission lines due to the presence of the $\delta$-functions. In order to find the linewidths one has to go beyond the kinetic equation approach and take into account the non-diagonal elements of the complete density matrix of the system $\varrho_{m,\mu;m',\mu'}$. It satisfies the following equation:
\begin{multline}
\label{eq:density}
\frac{\mathrm d \varrho}{\mathrm dt} = \frac{\mathrm i}{\hbar}[\varrho,\mathcal H] + \frac{\Gamma_{\rm C}}{2}(2c\varrho c^\dag - c^\dag c\varrho - \varrho c^\dag c) + \\
\sum_i \frac{\Gamma_{{\rm X},i}}{2}(2b_i^{\vphantom{\dag}}\varrho b_i^\dag - b_i^\dag b_i^{\vphantom{\dag}}\varrho - \varrho b_i^\dag b_i^{\vphantom{\dag}}) +\\
\sum_i \frac{W_i}{2}(2b_i^\dag\varrho b_i^{\vphantom{\dag}} - b_i^{\vphantom{\dag}} b_i^\dag\varrho - \varrho b_i^{\vphantom{\dag}} b_i^\dag).
\end{multline}
Here the first term describes the Hamiltonian-driven dynamics of the coupled quantum dots and cavity system, second term describes the photon escape through the cavity mirrors, last two terms stand for the non-radiative exciton decay and non-resonant exciton pumping, respectively. 
We use the general expression for the emission spectrum \cite{Carmichael}
\begin{equation}
\label{eq:general}
I(\omega) \propto \Re{\int_0^{\infty} \mathrm d t \mathrm e^{\mathrm i \omega t}\langle c^\dag(0) c(t)\rangle},
\end{equation}
where $c(t)$ and $c^\dag(t)$ are the time-dependent field operators for the cavity mode, the angular brackets denote both quantum mechanical and statistical averaging and express the correlation function of the photonic mode via the density matrix as
\begin{multline}
\label{eq:general1}
I(\omega) \propto \sum_{m,\mu,\mu'} f_{m,\mu} c_{m-1,\mu';m,\mu}^* \sum_{m_1,\mu_1,\mu_1'} c_{m_1-1,\mu'_1;m_1,\mu_1} \times \\
\tilde{\varrho}^{(m,\mu,\mu')}_{m_1,\mu_1;m_1-1\mu_1'}(\omega)\:.
\end{multline}
Here
\[
\tilde{\varrho}^{(m,\mu,\mu')}_{m_1,\mu_1;m_1-1\mu_1'}(\omega)=\Re{\int_0^{\infty} \mathrm d t \mathrm e^{\mathrm i \omega t}\varrho^{(m,\mu,\mu')}_{m_1,\mu_1;m_1-1\mu_1'}}(t),
\]
where $\varrho^{(m,\mu,\mu')}_{m_1,\mu_1;m_1-1\mu_1'}(t)$ are the time-dependent solutions of the total density matrix equation \eqref{eq:density}.
The upper subscripts $m,\mu,\mu'$ in Eq.~(\ref{eq:general1})  label the manifold, populated at $t=0$, so the initial conditions read
\begin{equation}\label{eq:initial}
\varrho^{(m,\mu,\mu')}_{m_1,\mu_1;m_1-1\mu_1'}(0)=\delta_{m,m_1}\delta_{\mu,\mu_1}\delta_{\mu',\mu_1'}.
\end{equation}
Thus, each term in the sum \eqref{eq:general1} with given indices $m$ and $\mu$ is proportional to the stationary population $f_{m,\mu}$, similarly to Eq.~(\ref{eq:Fermi}). In derivation of Eq.~\eqref{eq:general1} we took into account that, in the strong coupling limit, the stationary density matrix $\varrho^{\rm st}$ is diagonal in the basis of the manifolds, 
and can be found from the kinetic equation \eqref{eq:kinetic}, $\varrho^{\rm st}_{m,\mu;m',\mu'}=\delta_{m,m'}\delta_{\mu,\mu'}f_{m,\mu}$. 

 Eq.~\eqref{eq:general1} is the particular case of the more general expression presented in Ref.~\onlinecite{laussy2009} which is valid for the arbitrary coupling strength. In our case, the strong coupling condition results in the substantial reduction of the computational costs since much smaller number of the non-diagonal elements of the density matrix are to be calculated.
 If one neglects dissipative processes completely, then 
\[
\varrho^{(m,\mu,\mu')}_{m_1,\mu_1;m_1-1\mu_1'} = \e^{-\i ( E_{m,\mu}-E_{m-1,\mu'} )t/\hbar}\delta_{m,m_1}\delta_{\mu,\mu_1}\delta_{\mu',\mu_1'},
\]
and Eq.~\eqref{eq:general1} reduces to Eq.~\eqref{eq:Fermi}.


\section{Single quantum dot}\label{sec:1QD}
In this section we analyze emission spectra  of a microcavity with single quantum dot. Our calculation technique is tested in well-studied case of zero detuning between exciton and the photon mode\cite{laussy2009,savelyev2009,2010arXiv1005.2318Q} $(\omega_{\rm C}=\omega_{\rm X})$ and then applied to a general case, $\omega_{\rm C}\ne \omega_{\rm X}$.
 \begin{figure}[hptb]
 \begin{center}
  \includegraphics[width=0.35\textwidth]{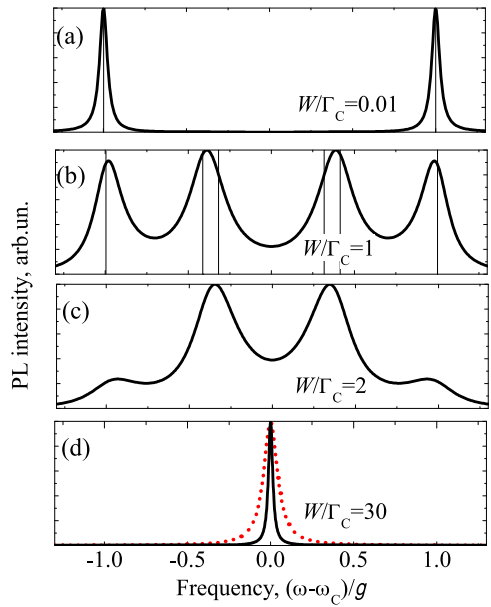}
 \caption{Photoluminescence spectra for $N=1$ quantum dot in resonance with the cavity mode for different pumping rates $W$.
 Panels (a)-(e) correspond to  $W/\Gamma_{\rm C}=0.01,1,2,30$, respectively, and are calculated for $\omega_{\rm X}=\omega_{\rm C},\Gamma_{\rm C}/g=0.1,\Gamma_{\rm X}/g=0.01$. Energies of the transitions, mainly determining the spectra, are indicated in panels (a), (b) by vertical lines. Dotted curve in panel (a) shows the Lorentzian with half-width equal to $\Gamma_{\rm C}/2$, corresponding to empty cavity spectrum in the linear regime.
  Photon mode energy is used as the reference point, the spectra are normalized to their maximum values.}\label{fig:1}
 \end{center} 
 \end{figure}
\subsection{Resonant quantum dot}\label{sub:0}

Photoluminescence spectra for zero detuning, calculated  for different rates of incoherent pumping rate $W$, are shown in Fig.~\ref{fig:1}.
Variation of the ratio between pumping and decay rates controls 
the population of the Jaynes-Cummings states, Eq.~\eqref{eq:jc:states}. Radiative transitions between these states determine optical spectra. Several regimes of light emission are resolved for different pumping rates:

(a) Linear regime, $W\ll\Gamma_{\rm C}$, see Fig.~\ref{fig:1}a. The mean number of polaritons
$\langle m\rangle\sim W/\Gamma_{\rm C}$ is much less than unity, so that  only the manifolds with $m=0$ and $m=1$ are populated, $f_{0}\gg f_{1,\pm}$. Photon emission from two states with $m=1$ results in the  Rabi doublet in spectrum with peaks at  
\[E_{1,\pm}-E_0=\hbar\omega_{\rm C}\pm\hbar g\:.
\]
In these case photon and exciton modes can be treated as coupled harmonic oscillators since average number of photons and excitons in the system is very small and the statistics of these states becomes unimportant.

(b) Intermediate regime, $W\sim\Gamma_{\rm C}$. The mean number of polaritons is on order of unity and several lowest manifolds are populated. The emission energies correspond to the difference of level energies in all relevant neighboring manifolds, hence, the spectrum has a complex multipeak stucture. In particular, for the case of Fig.~\ref{fig:1}(b), where $W=\Gamma_{\rm C}$, the peaks correspond to emission from manifolds with $m=1\ldots 3$, with the energies indicated by vertical lines.

(c) High pumping regime, $W\gg \Gamma_{\rm C}$. The mean number of photons is much larger than one, ${\langle m\rangle\approx W/(2\Gamma_{\rm C})\gg 1}$. 
Here, only the optical transitions between the states Eq.~\eqref{eq:jc:states} with the same symmetry, i.e. $m,+ \to m-1,+$,  are important. The corresponding photon energies are
\begin{equation}\label{eq:g_eff}
 E_{m,\pm}-E_{m-1,\pm}\approx \hbar\omega_{\rm C}\pm \frac{\hbar g}{2\sqrt{m}}\:,
\end{equation}
resulting in a two-peak emission spectrum, see Fig.~\ref{fig:1}c.
Simple analytical expression
\begin{equation}\label{eq:I_an_low_pump}
 I(\omega)\propto \frac{1}{(\omega-\omega_0- g_{\rm eff})^2+\gamma^2}+
\frac{1}{(\omega-\omega_0+g_{\rm eff})^2+\gamma^2}\:,\end{equation}
well describes the spectrum, see Appendix~\ref{sec:A1} for details of derivation. The doublet \eqref{eq:I_an_low_pump} is characterized by 
effective Rabi splitting  $2g_{\rm eff}=g/\sqrt{\langle m\rangle}$ 
and peak half-width $\gamma=(W+\Gamma_{\rm C})/4$.
Hereafter we neglect exciton non-radiative decay as compared with the photon leakage through the mirrors.
We note, that although Eq.~(\ref{eq:I_an_low_pump}) has a simple form, it is determined by a lot of optical transitions Eq.~\eqref{eq:g_eff} with different $m$.
As shown in Appendix~\ref{sec:A1}, the interplay between these transitions
due to the fermionic nature of the exciton leads to the broadening of the peaks with respect to the linear regime, cf. Figs.~\ref{fig:1}a and \ref{fig:1}c. Moreover, the distance between the peaks is much smaller than in linear case, because
$g_{\rm eff}\ll g$. 
This effect allows  semiclassical interpretation, since the mean number of cavity photons is much larger than unity. The decrease of $g_{\rm eff}$ with pumping can be considered as quenching of the optical transition due to the saturation of two-level dot.\cite{Mukamel} In agreement with this concept, the average number of excitons is close to $1/2$, just like for a two-level system, interacting with strong (classical) electromagnetic field.\cite{panthell}

(d) The further increase of pumping rate leads to the the transition to lasing regime, which takes place when $W$ is equal to
\begin{equation}
\label{Wstar}
 W^*=2g^{2/3}\Gamma_{\rm C}^{1/3}\:.
\end{equation}
The lasing regime is manifested as single-peak spectrum, with half-width rapidly decreasing when pumping rate grows. The approximate analytical expression for spectrum reads
 \begin{equation}\label{eq:I_an_high_pump}
I(\omega)\propto \frac{1}{(\omega-\omega_{\rm C})^2+(g^2\Gamma_{\rm C}/W^2)^2}\:.
\end{equation}
In agreement with this result Fig.~\ref{fig:1}d demonstrates, that the emission spectrum becomes much narrower than in the linear case, cf. solid and dashed curves. The spectral width $\Gamma_{\rm C}(g/W)^2$ is also smaller than the effective Rabi splitting $2g_{\rm eff}$.

Further increase of the pumping results in the self-quenching of a single-dot laser regime,\cite{laussy2009,mu1992} and, correspondingly, violates the strong coupling condition. Indeed, if the pumping rate exceeds the splitting between the $+$ and $-$ states in the `actual' manifold, $W>\sqrt{\langle m\rangle} g$ or
\[
W>\frac{g^2}{\Gamma_C},
\]
the states $+$ and $-$ within the given manifold become ill-defined. It is worth noting that, in high finesse microcavities, with $g\gg \Gamma_{\rm C}$, the characteristic pumping rates separating different spectral regimes are considerably different, $\Gamma_{\rm C}\ll W^*\ll g^2/\Gamma_{\rm C}$. 
%
%
Provided that $W<g^2/\Gamma_C$ the strong coupling regime is maintained and our results are in the perfect agreement  with those obtained in Ref.~\onlinecite{laussy2009} by keeping all elements of the density matrix which provides the justification of calculation technique presented in Sec.~\ref{sec:model}.




 \begin{figure}[hptb]
\begin{center}
 \includegraphics[width=0.35\textwidth]{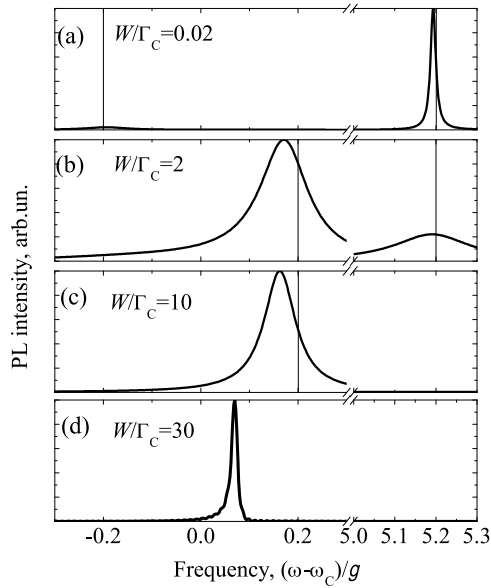}
\caption{Photoluminescence spectra for $N=1$ quantum dot detuned from the cavity resonance for different pumping rates $W$.
Panels (a)--(e) correspond to  $W/\Gamma_{\rm C}=0.02,2,10,30$, respectively, and are calculated for detuning $\omega_{\rm X}-\omega_{\rm C}=5g$. Other parameters are the same as for Fig.~\ref{fig:1}.
 Energies of the transitions, mainly determining the spectra, are indicated in panels (a)--(c) by vertical lines.  Photon mode energy is used as the reference point, the spectra are normalized to their maximum values.}\label{fig:2}
\end{center} 
\end{figure}

\subsection{Detuned quantum dot}\label{sub:det}

\begin{figure}[hptb]
\begin{center}
 \includegraphics[width=0.4\textwidth]{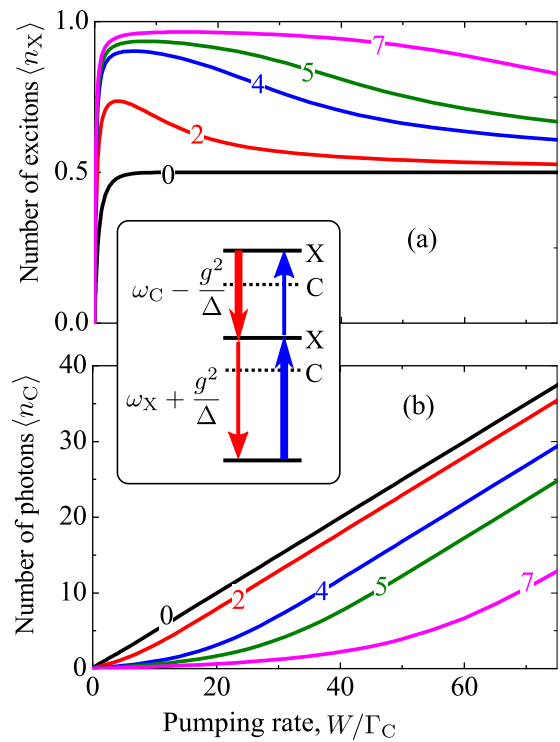}
\caption{(Color online) Average numbers of excitons (a)
and numbers of photons (b) as functions of the pumping rate, calculated for the values of detuning $(\omega_{\rm X}-\omega_{\rm C})/g=0,2,4,5,7$, indicated at each curve. Other parameters are the same as for Fig.~\ref{fig:1}. The inset schematically illustrates the three lowest manifolds of Jaynes-Cummings ladder.
Vertical arrows indicate radiative and non-radiative transitions between the exciton-like (\textrm{X}) states, governing the kinetics of the system. 
} \label{fig:3}
\end{center}
\end{figure}

Now we turn to the case when the exciton energy is detuned from the photon mode. Figure~\ref{fig:2} presents 
emission spectra calculated at different pumping rates $W$ for large detuning $\Delta\equiv\omega_{\rm X}- \omega_{\rm C}=5g$, and Fig.~\ref{fig:3} shows how the mean numbers of  excitons $\langle n_{\rm X}\rangle $ and photons $\langle n_{\rm C}\rangle $ depend on pumping rate for different values of $\Delta $. 
The figures reveal nontrivial behaviour of spectra and particle statistics when $W$ changes. To understand it we  analyze the  eigenstates of the Hamiltonian \eqref{eq:Hamiltonian}.

Two $m$-particle states can be considered 
for sufficiently large detuning $(g\sqrt{m} \ll \Delta )$ 
as exciton-like (X) and cavity photon-like state (C). Their energies in the second order of perturbation theory in $g$ are given by
\begin{subequations}\label{eq:E_exc}
\begin{equation}
E^{({\rm C})}_m=m\hbar \omega_{C}- m g^2/\Delta,
\end{equation}
\begin{equation}
 E^{({\rm X})}_m=(m-1)\hbar \omega_{\rm C}+\hbar\omega_{\rm X}+m g^2/\Delta,\quad m=1,2\ldots.
\end{equation}
\end{subequations}
Only the exciton-like states  
\begin{equation}\label{eq:exc}
 |m{\rm X}\rangle= |m-1, 1\rangle +\frac{g\sqrt{m}}{\Delta} |m, 0\rangle,\quad m=1,2,\ldots 
\end{equation}
are populated by pumping to the quantum dot and they determine the emission spectrum.
The first term in \eqref{eq:exc} corresponds to occupied dot and $m-1$ photons, while the second decribes small admixture of the state with empty dot and $m$ photons.

Let us consider which eigenstates determine the emission spectra for different pumping.
For very small pumping rates, corresponding to Fig.~\ref{fig:2}a, only the  manifold with $m=1$ is populated. The spectrum consists of the peak close to the exciton energy $E_1^{\rm X}$ and very weak peak at the energy $E_1^{\rm C}$ of the almost empty photon-like state. Such spectrum is typical for linear in pumping regime at large detuning.\cite{JETP2009}

Interestingly, the state with one exciton,  $|1\mathrm X \rangle$, becomes significantly populated already at small pumpings. Indeed, the pumping rate to this state is just
 \[ W^{\rm X}_{0\to |1{\rm X}\rangle }=W,\]
while its decay rate (determined by the photon escape from the cavity) is small:
\begin{equation}
 W^{\rm C}_{|1{\rm X}\rangle \to 0}=\Gamma_{\rm C}\times \left(\frac{g}{\Delta}\right)^2,
\end{equation}
since photonic fraction is almost negligible in this state. At $W\gtrsim W^{\rm C}_{|1{\rm X}\rangle \to 0}$ the dot is already almost completely occupied while the cavity is still almost empty, $\langle n_{\rm C}\rangle \sim (g/\Delta)^2\ll \langle n_{\rm X}\rangle \approx 1$, cf. Figs.~\ref{fig:3}, a and b. 

By contrast, the pumping to the second state $|2{\rm X}\rangle$ is inefficient,
\begin{equation*}
 W^{\rm X}_{|1{\rm X}\rangle \to 2{\rm X}\rangle}=W\times \left(\frac{g}{\Delta}\right)^2\ll W\:,
\end{equation*}
while its radiative decay rate is high, 
\[
W^{\rm C}_{|2{\rm X}\rangle \to |1{\rm X}\rangle}\approx \Gamma_C.
\]
 Consequently, to populate second state Eq.~\eqref{eq:exc} high pumping is required, contrary to the first state, see the inset to Fig.~\ref{fig:3}. 
Interestingly, that relatively high value of radiative rate $W^{\rm C}_{|2{\rm X}\rangle \to |1{\rm X}\rangle}$ means that  even for negligible population of the state
$|2{\rm X}\rangle$ its contribution to the emission spectrum can be already important. The frequency of the corresponding peak  is close to the photon mode, $E_{2}^{\rm X}-E_{1}^{\rm X}=\hbar\omega_{\rm C}+\hbar g^2/\Delta$, see Fig.~\ref{fig:2}b. This peak dominates in the spectrum when the pumping rate exceeds~$\Gamma_{\rm C}$.

Increase of pumping rate up to
\begin{equation}
 W\sim \Gamma_{\rm C}\times \left(\frac{\Delta}{g}\right)^2,
\end{equation}
makes population of the second state $|2{\rm X}\rangle$  considerable (in this case $W^{\rm X}_{|1{\rm X}\rangle \to 2{\rm X}\rangle}\sim W^{\rm C}_{|2{\rm X}\rangle\to |1{\rm X}\rangle}$).
At such pumping  average exciton and photon numbers are both on order of unity, as is demonstrated by Fig.~\ref{fig:3}.
The further increase of the pumping rate leads to the population of higher states \eqref{eq:exc} with $m\ge 3$. For higher $m$ the exciton fraction 
in states \eqref{eq:exc} is smaller so that the number of excitons slowly decreases and tends to $1/2$, as in case of  zero detuning. Thus, the exciton population has a plateau at $(g/\Delta)^2\lesssim W/\Gamma_{\rm C}\lesssim (\Delta/g)^2$. This plateau becomes more prominent for higher values of detuning, see Fig.~\ref{fig:3}a.

The ratio between the emission and pumping rates for $m\ge 3$ manifolds remains the same as for transition $|2{\rm X}\rangle \to |1{\rm X}\rangle$, so the photon number increases linearly with pumping, see  Fig.~\ref{fig:3}b. The peak frequency remains approximately equal to $\omega_{\rm C}+g^2/\Delta$ because the states $E_m^{\rm X}$ are almost equidistant.
At even higher pumping rates, the detuning becomes smaller as compared to the effective exciton-photon coupling $g\sqrt{n_{\rm C}}$ so that our analysis presented in Sec.~\ref{sub:0} for high powers at zero detuning becomes applicable. Increase of pumping leads to the transition into lasing regime with narrowing spectrum, cf. Figs.~\ref{fig:2}c and \ref{fig:2}d.


\section{Mesoscopic effects in emission}\label{sec:mes}
In this Section we consider the microcavity where $N>1$ quantum dots are close to the photonic mode resonance.
First we ignore the detuning between the excitonic and photonic energies and then take  detuning into account.

In case when all $N$ quantum dots are the same
($\omega_{\rm X,i}=\omega_{\rm C}$, $\Gamma_{{\rm X},i}\equiv \Gamma_{\rm X}$, $W_i\equiv W$),  emission spectra at low pumping  ($NW\ll \Gamma_{\rm C}$) are determined by the symmetrical, superradiant mode, where all the excitons oscillate in phase.\cite{JETP2009,Dicke1954}
Superradiant mode belongs to the manifold with $m=1$ particle, which consists of $N+1$
states in total, 
\begin{align}
 E_{1,\rm SR,\pm}&=\hbar\omega_{\rm C}\pm \sqrt{N}\hbar g,\label{eq:E_SR}\\
E_{1,\rm dark}&=\hbar\omega_{\rm C},\label{eq:dark}
\end{align}
$N-1$ of which, given by Eq. (\ref{eq:dark}), are dark. Thus, the emission spectrum has a two-peak structure with the splitting between the peaks enhanced by the factor $\sqrt{N}$.
Increase of the pumping rate leads to the population of the higher manifolds with 
$m\ge 2$. Extra peaks arise in the spectrum due to the radiative transitions between the manifolds $m=2$ and $m=1$ since dark states of the first manifold may serve as final states for the transitions from the second one.
If the number of dots is larger than unity, the transitions between second and first manifold become manifested at substantially lower powers as compared with a single dot case.
This effect is related to the dark states Eq.~\eqref{eq:dark}. While the pumping rates to the superradiant and dark states are the same, the total decay rates $D_{1,\rm dark}$ of the dark states in the first manifold are much smaller since photon emission is not possible. For instance, if $\Gamma_{\rm C}$  is negligible,  dark states decay only via pumping  to the second manifold, $W^{\rm X}_{|1,{\rm dark}\rangle \to 2,\mu}$. Consequently, they are already populated at small pumping and act as a reservoir for the pumping of the second manifold. It is similar to the role of the long-living state $|1{\rm X}\rangle$ with long radiative lifetime in pumping mechanism  considered in the previous Section, when the single dot detuned from the photonic mode was analyzed.

\begin{figure}[hptb]
\begin{center}
 \includegraphics[width=0.48\textwidth]{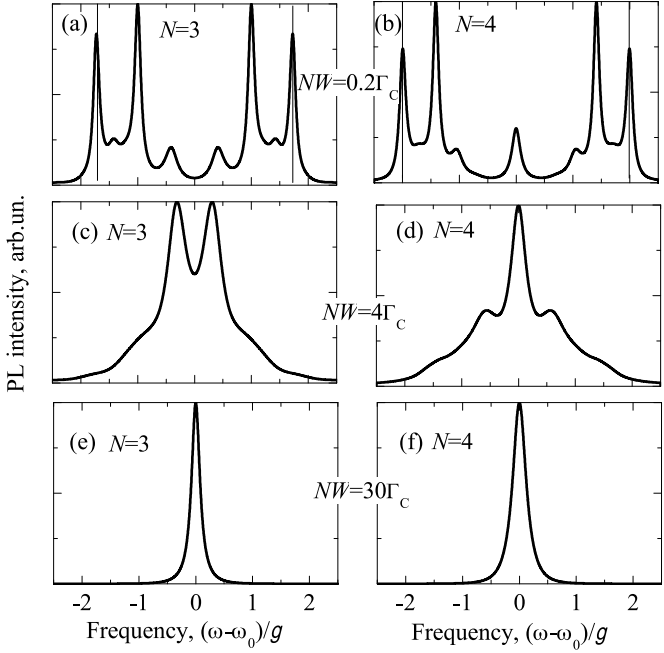}
\caption{Photoluminescence spectra for $N=3$ (panels a, c, e) and $N=4$ (panels b, d, f) quantum dots for different pumping rates $W$.
Panels (a) and (b), (c) and (d), (e) and (f) were calculated for  $NW/\Gamma_{\rm C}=0.2,4,30$, respectively. Other parameters are the same as for Fig.~\ref{fig:1}.
 Energies of the transitions corresponding to the superradiant mode are indicated at panels (a) and (b)  by vertical lines.  Photon mode energy is used as the reference point, the spectra are normalized to their maximum values. 
}\label{fig:5}
\end{center} 
\end{figure}

It turns out that, the superradiant regime is destroyed by the relatively weak pumping. The emission spectra for $N=3$ and $N=4$ quantum dots presented in Figs.~\ref{fig:5}a,~\ref{fig:5}b at small pumping rate $NW=0.2\Gamma_{\rm C}$ already show complex multipeak structure arising from the population of the higher manifolds.

 \begin{figure}[b!]
\begin{center}
 \includegraphics[width=0.35\textwidth]{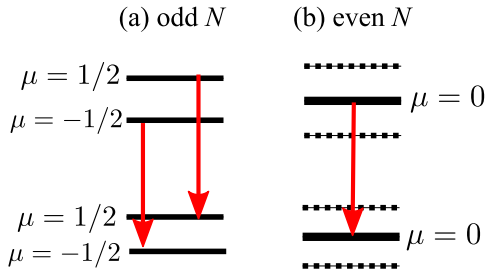}
\caption{Scheme of the radiative transitions for odd (a) and even (b) numbers of quantum dots $N$. }\label{fig:4}
\end{center} 
\end{figure}

Let us now  focus on the regime of large pumping. For  $W\gg \Gamma_{\rm C}$ the typical number of cavity photons is large, so the electromagnetic field can be considered classicaly.
Then the  Hamiltonian \eqref{eq:Hamiltonian} can be considerably   simplified by replacement of the photon creation and annihilation operators $c^\dag$ and $c$  by their
expectation values. For $m$-th manifold  expectation values are approximately equal to $\sqrt{m}$. Hence, the Hamiltonian \eqref{eq:Hamiltonian} has the form
\begin{equation}\label{eq:H2}
 H_m\approx 
m\hbar\omega_{\rm C}+\sqrt{m}\hbar g\sum_i (b_i^{\vphantom{\dag}}+ b_i^\dag)\:,
\end{equation}
including only excitonic operators. To diagonalize it we notice, that the two-level dot behaves like the spin $1/2$.
In particular, $
b_i^{\vphantom{\dag}}+ b_i^\dag$ can be presented as $2s_z^{(i)}
$, where $s_z^{(i)}$ is the operator of $z$ projection of spin $1/2$ with eigenvalues $\pm 1/2$.
After summation over all $N$ dots we obtain the operator of total spin  projection. Its eigenvalues are equidistantly distributed between
$-N/2$ and $N/2$, so the spectrum of Hamiltonian \eqref{eq:H2} reads
\begin{equation}\label{eq:spin}
 E_{m,\mu}=\hbar m\omega_{\rm C}+2\mu\sqrt{m}\hbar g,
\end{equation}
\[
\mu=-\frac{N}{2},-\frac{N}{2}+1,\ldots,\frac{N}{2}-1,\frac{N}{2}\:.
\]
Each state $\mu$ corresponds to $\mu+N/2$ dots with ``spin up'' and $-\mu+N/2$ dots
with ``spin down''. State degeneracy equals to $C_{N}^{\mu+N/2}$ and the total number of states \label{eq:spin} is $2^N$. 
The radiative  transitions between manifolds $m$ and $m-1$ conserve $\mu$ in used approximation. Decay and pumping rates are $\mu$-independent. 
Thus, emission spectrum is determined by the states with the smallest $|\mu|$, which have the highest degeneracy.
It is important, that the smallest value of $|\mu|$ equals to either $1/2$ or $0$
depending on the parity of the number of dots, so
\begin{gather}
 E_{m,\pm 1/2}-E_{m-1,\pm 1/2}\approx \hbar\omega_{\rm C}\pm \frac{\hbar g}{2\sqrt{m}} \:\:\:\text{(odd $N$)},\\
E_{m,0}-E_{m-1,0}\approx \hbar\omega_{\rm C}\:\:\:\text{(even $N$)}\:.
\end{gather}
These two cases are schematically illustrated in Figs.~\ref{fig:4}a and \ref{fig:4}b. We conclude,  that for odd number of dots and high pumping rates [$\Gamma_{\rm C}\ll NW\ll W^*$ where $W^*$ is given by Eq.~\eqref{Wstar}] the emission spectrum has a two-peak shape, like for $N=1$, while of even $N$ it consists of single peak at $\omega=\omega_{\rm C}$. 
 \begin{figure}[b!]
\begin{center}
 \includegraphics[width=0.4\textwidth]{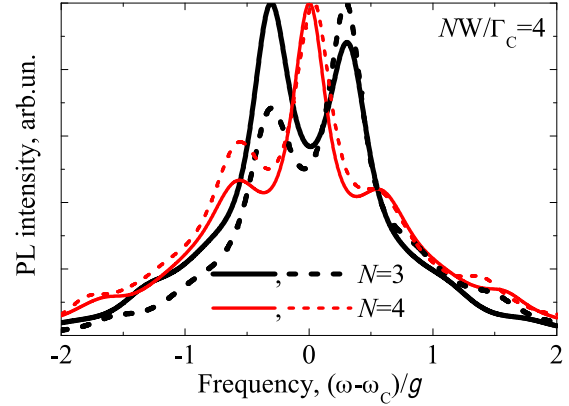}
\caption{(Color online) Photoluminescence spectra calculated taking into account inhomogeneous broadening.
Thick (black) and thin (red) curves correspond to $N=3$ and $N=4$ dots, respectively.
Solid and dashed curves were calculated for different particular realizations of random spread of excitonic energies. The value of pumping rate was $NW=4\Gamma_{\rm C}$.
Arithmetic mean values of the detuning are $\langle|\omega_{{\rm X},i}-\omega_{\rm C}|\rangle/g \approx 0.4,0.5,0.6,1$ for black solid, black dashed, red solid, red dotted curves, respectively. 
 Other parameters are the same as for Fig.~\ref{fig:1}. Photon mode energy is used as the reference point, the spectra are normalized to their maximum values.
}\label{fig:6}
\end{center} 
\end{figure}

Our above analysis is fully confirmed by the results of numerical calculation presented on Fig.~\ref{fig:5}. Panels (a)-(f) show the dependence of spectra for $N=3$ and $N=4$ dots on pumping rate.
As we mentioned above, the superradiant regime is already destroyed
at relatively low pumping $NW=0.2\Gamma_{\rm C}$  since the dark states increase the pumping efficiency to the second manifold, see panels (a), (b).
At relatively high pumping $NW=4\Gamma_{\rm C}$, shown in  panels  (c) and (d), the distance between peaks for both $N=3$ and $N=4$ becomes smaller due to saturation of oscillator strength, while individual peaks are wider, just like for a single dot. However, the spectral shape strongly depends on the parity of $N$, the doublet corresponds to $N=3$ and singlet corresponds to $N=4$.
Finally, at high pumping rates  the system is in the lasing regime, see Figs.~\ref{fig:5}e and \ref{fig:5}f. Its spectrum consists of single peak which width decreases with the pumping power.

We note, that the Hamiltonian \eqref{eq:Hamiltonian} has been already
 extensively studied for cavity with $N\gg 1$ atoms. Such a system is well described by the Tavis-Cummings model.\cite{tavis1968,tavis1969} The specific feature of the semiconductor cavity is that the number of quantum dots strongly coupled with photonic mode cannot be very large, $N\sim 1\ldots 10$. Parity-sensitive mesoscopic effects 
demonstrated above become then important. 

Moreover, the inevitable disorder leading to the spread of the excitonic resonant frequencies $\omega_{{\rm X},i}$ makes semiconductor systems particularly different from the atomic cavities. In order to study the disorder effect we calculated the emission spectra of three and four quantum dots in a microcavity for the different realizations of the exciton resonance frequencies, see Fig.~\ref{fig:6}. 
For all four curves in Fig.~\ref{fig:6}  the mean arithmetic values of the detuning $(1/N)\sum_{i=1}^N|\omega_{{\rm X},i}-\omega_{\rm C}|$ were in the order of the light-matter coupling constant~$g$.
 Although the spectra are modified by disorder, the
doublet ($N=4$) can be still well distinguished  from the singlet ($N=3$).
Thus, we conclude that the parity-dependent mesoscopic effects in emission spectra are relatively stable against the disorder.


\section{Conclusions}\label{sec:concl}

To summarize, we have developed a kinetic theory of the 
non-linear emission of
quantum dots embedded into the semiconductor microcavity. We considered the case of the strong coupling where the light-matter interaction constant is larger than the damping rates of the cavity and exciton, governed by the photon escape through the mirrors and exciton non-radiative decay, respectively. In this case the eigenstates of the system belonging to the manifolds with the total particle number $m$ are well defined so that the emission spectra can be found from the Fermi golden rule. The populations of the coupled photon-exciton states are determined by the kinetic equation which takes into account both pumping and decay processes. The linewidths of the emission spectra are found as corrections to the kinetic equations.

Our method was tested in the case of a single quantum dot in a microcavity being resonant with the photonic mode. In the strong coupling regime we reproduced the results of Ref.~\onlinecite{laussy2009}. The transition from the linear regime with Rabi doublet via multipeak spectrum to the narrow emission line being the characteristic of the laser is demonstrated.

We have studied in detail the case of the single dot detuned from the cavity mode. At low pumping rates the emission is dominated by the quantum dot line. An increase of the pumping results in the population of the second manifold and the emission line is close to the cavity position. Even higher pumping results in the line narrowing and transition to the lasing regime similarly to the case of the resonant quantum dot.  

The eigenmodes of the interacting cavity-quantum dots system can be found analytically in the case of large photon number where the cavity field is classical and the light-matter coupling can be formulated in the terms of angular momentum. In the situation where several dots are embedded into the microcavity the emission spectra are shown to demonstrate parity effect: at relatively strong pumping there are one or two peaks in the spectrum depending on whether the dot number is even or odd, respectively. This parity effect is relatively robust to the spread of exciton resonance energies caused, e.g., by the disorder.

To conclude, the non-linear emission spectra of the quantum dots in a cavity provide important information both about light-matter interaction and about the quantum dot ensemble.

\acknowledgments

The authors thank  F.P. Laussy and E. del Valle for useful discussions.
The financial support of the RFBR, Programs of the RAS, FASI, President grant for young scientists and the ``Dynasty'' Foundation -- ICFPM is
gratefully acknowledged.

\appendix 

\section{Analytical solution for single QD in microcavity}
 \label{sec:A1}
In this Appendix we present analytical solutions for polariton distribution function and emission spectrum of single quantum dot, strongly coupled with microcavity photon mode. The detuning is neglected, $\omega_{\rm C}=\omega_{\rm X}$.

We  consider relatively high pumping, $W\gg \Gamma_{\rm C}$, so that the spectrum is determined by high manifolds, $m\gg 1$. Moreover, the strong coupling regime is assumed, i.e. $g\gg \Gamma_{\rm C}$, $\sqrt{m}g \gg W$, the exciton decay is neglected.
In this case only states with the same symmetry are coupled by the photon annihilation operator:
\begin{equation}\label{eq:c_diag}
c_{m-1,\mu;m\mu'}\approx\sqrt{m}\delta_{\mu,\mu'}\:.
\end{equation}
Clearly, under these assumptions states $\mu=+$ and $\mu=-$ are equally populated. First we need to determine the stationary distribution function of polaritons $f_{m}=f_{m,+}+f_{m,-}=2f_{m,+}=2f_{m,-}$. Under the assumption \eqref{eq:c_diag} kinetic Eq.~\eqref{eq:kinetic} reads
\begin{equation}\label{eq:kin1}
 \frac{\mathrm df_m}{\mathrm dt}=-\Gamma_{\rm C}mf_m+\Gamma_{\rm C}(m+1)f_{m+1}+\frac{W}{2}(f_{m-1}-f_{m}).
\end{equation}
Since typical values of $m$ are high, the discrete function $f_m$ can be replaced by continuous distribution $f(m)$. Finite-difference equation \eqref{eq:kin1} then, at steady state, becomes differential equation
\begin{equation}\label{eq:kin2}
  (m+\langle m\rangle)\frac{\mathrm d^2f}{\mathrm dm^2}+2(m-\langle m\rangle)\frac{\mathrm df}{\mathrm dm} +2f=0\:,
\end{equation}
where
$
\langle m\rangle=W/(2\Gamma_{\rm C})\:.
$
The solution of Eq.~\eqref{eq:kin2}, describing the distribution function, reads
\begin{equation}
 f(m)\propto (m+\langle m\rangle)^{4\langle m\rangle+1}\e^{-2m}\:,
\end{equation}
where the normalization constant to be determined from Eq.~\eqref{eq:norm} is omitted.
Thus, for high pumping, when ${\langle m\rangle \gg 1}$, the distrubution function has Gaussian shape,
\begin{equation}\label{eq:Gauss}
  f(m)\propto \exp\Bigl(-\cfrac{(m-\langle m\rangle)^2}{2\langle m\rangle}\Bigr)\:,
\end{equation}
with $\langle m\rangle$ being the average number of polaritons.

Now we proceed to the calculation of emission spectrum. In the strong coupling regime the only relevant off-diagonal components of the density matrix are those with $\mu=\mu'$, which can be presented as
$
\varrho_{m,\pm ;m-1,\pm}=(a_m\pm \i b_m)\exp(-\i\omega_{\rm C}t)\:,
$
where $a_m$ and $b_m$  are real coefficients.
Eq.~\eqref{eq:density} leads to
\begin{gather}
\dot a_m(t)=g_mb_m(t)+\frac{W}{2}[a_{m-1}(t)-a_m(t)]+(\bm\Gamma_{\rm C} a)_m\:\nonumber,\\\label{eq:ab1}
\dot b_m(t)=-g_ma_m(t)-\frac{W}{2}b_m(t)+(\bm\Gamma_{\rm C} b)_m \:,
\end{gather}
where
$g_m\equiv g(\sqrt{m} -\sqrt{m-1})$  and  the term
\begin{equation}
 (\bm\Gamma_{\rm C} d)_m=\Gamma_{\rm C}\left[\sqrt{m(m+1)}d_{m+1}-\Bigl(m-\frac{1}{2}\Bigr)d_m\right],d=a,b\:,
\end{equation}
describes the cavity decay.
Initial conditions \eqref{eq:initial} are 
\begin{equation}\label{eq:initial1}
 a_m(0)=\sqrt{m}f_m,\quad b_m(0)=0
\end{equation}
 and emission spectrum \eqref{eq:general1} is given by
\begin{equation}
 I(\omega)\propto\sum\limits_m\sqrt{m}\int_0^\infty\cos[(\omega-\omega_{\rm C}) t]a_m(t)dt\:.
\end{equation}
 Similarly to kinetic equation, system \eqref{eq:ab1} can be solved efficiently by introducing continuous functions $a(m),b(m)$, which satisfy differential equations
\begin{align}
\frac{\mathrm da}{\mathrm dt}&=g_{m} b-\frac{P_{\rm X}}{2}\frac{\mathrm  da}{\mathrm  dm}+\Gamma_{\rm C}\left[a+m\frac{\mathrm da}{\mathrm  dm}\right]\label{eq:ab2},\\
\frac{\mathrm db}{\mathrm dt}&=-g_{m} a-\frac{P_{\rm X}}{2}b+\Gamma_{\rm C}\left[b+m\frac{\mathrm  db}{\mathrm  dm}\right]\nonumber\:.
\end{align}
Keeping in mind initial conditions \eqref{eq:initial1} we seek the solutions of \eqref{eq:ab2} in the form
\begin{equation}\label{eq:ansatz}
 a(m)=\sqrt{m}f(m) A(t),\quad b(m)=\sqrt{m}f(m) B(t)\:.
\end{equation}
To proceed further we integrate \eqref{eq:ab2} over $m$ using identities
\begin{gather}
\int dm a(m)\approx \sqrt{\langle m\rangle} A, \int dm \sqrt{m}a(m)\approx\langle m\rangle A\:\label{eq:id1},\\
\int dm m^{3/2}\frac{\mathrm da}{\mathrm  dm}\approx-\frac{3\langle m\rangle}{2}A,\:\int dm \sqrt{m}\frac{\mathrm da}{\mathrm  dm}\approx \frac{A}{2},\nonumber\\
\int dm (\sqrt{m}-\sqrt{m-1})a(m) \approx-\frac{A}{2}\nonumber,
\end{gather}
which follow from Eqs. \eqref{eq:Gauss}, \eqref{eq:ansatz} in the leading order over $\langle m\rangle$.
The functions $A(t)$, $B(t)$ satisfy the coupled differential equations
\begin{equation}\label{eq:AB}
 \frac{\mathrm dA}{\mathrm dt}=g_{\rm eff}B,\:\:
\frac{\mathrm dB}{\mathrm dt}=-g_{\rm eff}A-2\gamma B
\end{equation}
with initial conditions $A(0)=1$, $B(0)=0$, where $g_{\rm eff}\approx g/(2\sqrt{\langle m\rangle})$ and $\gamma=(W+\Gamma_{\rm C})/4$.
The solution reads
\begin{equation}
 A(t)=\exp(-\gamma t)\left(\cos \Omega t+\frac{\gamma}{\Omega}\sin \Omega t\right),
\end{equation}
with $\Omega=\sqrt{g_{\rm eff}^2-\gamma^2}$. Finally, the emission spectrum is obtained by Fourier transformation, $
 I(\omega)\propto \langle m\rangle \int_0^\infty\cos[(\omega-\omega_{\rm C}) t] A(t)dt
$\:.

At a threshold value of the pumping,
$W=W^*=2g^{2/3}\Gamma_{\rm C}^{1/3}$ the  $\Omega$ turns to zero.
 The spectrum
has  different form, depending on the relation between the pumping rate $W$ and $W^*$.
For $W<W^*$ the value of $\Omega$ is real and the spectrum has two-peak shape
\begin{equation}\label{eq:an_low}
 I(\omega)\propto\frac{2\Omega-\omega+\omega_{\rm C}}{(\omega-\omega_{\rm C}-\Omega)^2+\gamma^2}+
\frac{2\Omega+\omega-\omega_{\rm C}}{(\omega-\omega_{\rm C}+\Omega)^2+\gamma^2}\:.
\end{equation}
The peak width is increasing with pumping.
At relatively low $(\Gamma_{\rm C}\ll W\ll W^*)$  pumping rates \eqref{eq:an_low} can be simplified to
Eq.~\eqref{eq:I_an_low_pump}.
Above the threshold  the spectrum has single peak shape
\begin{equation}\label{eq:an_high}
  I(\omega)\propto \frac{1}{(\omega-\omega_{\rm C})^2+(|\Omega|-\gamma)^2}+
 \frac{1}{(\omega-\omega_{\rm C})^2+(|\Omega|+\gamma)^2}\:.
\end{equation}
Only the first term in Eq.~\eqref{eq:an_low} is of interest at high pumping ($W\gg W^*$), leading  to the peak \eqref{eq:I_an_high_pump}, narrowing when $W$ increases. We remind that our theory is valid at $W\ll g^2/\Gamma_{\rm C}\sim W^* \times (g/\Gamma_{\rm C})^{4/3}$. At very high pumping rate the laser self-quenching  takes  place.\cite{mu1992,laussy2009}
\section{Effects of the blueshift, random sources approach} \label{sec:A2}

In this Appendix we consider the effects of exciton-exciton interactions on the emission spectra of quantum dots embedded in the microcavities. In order to elucidate the role of interactions we use classical model of Ref.~\onlinecite{JETP2009} and treat excitons and photons as classical oscillators (quantum treatment for the model case of interactions is presented in Ref.~\onlinecite{DelValle2008} and for the lasing regime in Ref.~\onlinecite{Ritter2010}). These oscillators are conveniently described by the dimensionless exciton polarizations $P_i = \langle b_i\rangle$ ($i=1,\ldots, n$) and dimensionless electric field amplitude in the cavity $E=\langle c\rangle$. Corresponding linear equations of motion can be derived from Hamiltonian Eq.~\eqref{eq:Hamiltonian} by considering Heisenberg equations of motion for the excitonic and photonic operators, taking average of the Heisenberg equations and neglecting high order correlators of exciton fields. So far, exciton-exciton interactions were disregarded. In the semiclassical approach they can be introduced as non-linear terms in the coupled oscillators model:
\begin{subequations}
\label{nonlin}

\begin{gather}
\label{eq:Et}
\frac{dE}{dt} = -\left(\mathrm i \omega_{\rm C} + \frac{\Gamma_{\rm C}}{2}\right) E + g \sum\limits_{i}P_i,\\
\label{eq:Pt}
\begin{split}\frac{d P_i}{dt} = -\left(\mathrm i \omega_{X,i} P_i + \frac{\Gamma_{{\rm X},i}}{2}  +  \sum_{j} \alpha_{ji} |P_j|^2 \right) P_i 
+\\\hspace{2cm} g E + w_i(t), \quad i=1,\ldots, N.
\end{split}
\end{gather}
\end{subequations}
Equation for the cavity field remains linear while the equations for the polarizations, Eq.~\eqref{eq:Pt}, acquire anharmonic contributions described by the real coefficients $\alpha_{ji}$ which determine energy shift of $i$th exciton due to the interaction with $j$th exciton. Rigorous derivation of $P^3$ terms and discussion of their microscopic origin is out of the scope of this paper. Terms $w_i(t)$ in Eq.~\eqref{eq:Pt} are the random forces which determine exciton generation in quantum dots.~\cite{JETP2009}

In order to make the analysis more transparent we, instead of studying Eqs.~\eqref{nonlin} in their complexity focus on the single exciton in one quantum dot and take into account simplest possible interaction in the form $\alpha |P|^2P$. 
\newcommand{\av}[1]{\left\langle#1\right\rangle}
The frequency spectrum  
of such an oscillator driven by the white-noise random force
\[
\langle w(t)w(t') \rangle = W \delta(t-t'),
\]
where the brackets denote averaging over the different realizations of the pumping, 
is related to the joint probability
$\mathcal F(P,t;P',t')$, which can be rigorously determined by Fokker-Planck equation technique.\cite{Risken_1989,gitterman}

Similarly to the Eq.~\eqref{eq:general} in the quantum case,
the spectrum reads 
\begin{gather}
 \av{|P(\omega)|^2} \propto \Re\int_0^\infty dt \av{P(t)P^*(0)}\e^{\i\omega t} 
\end{gather}
where the correlator $\av{P(t)P^*(0)}$ is given by
\begin{equation}
 \av{P(t)P^*(0)}=\iint  P P'^*\mathcal F(P,t;P',0) d^2P d^2P',
\end{equation}
and $d^2P\equiv d\Re P d\Im P$.
Joint probability can be expressed as\cite{Risken_1989}
\begin{equation}
\mathcal F(P,t;P',0)=\mathcal G(P,t|P',0) \mathcal F(P')  
\end{equation}
via the conditional probability 
$\mathcal G(P,t|P',0)$ and the stationary distribution function
 $\mathcal F(P')\equiv \lim_{t\to \infty} \mathcal G(P',t|P'',0)$.
The conditional probability satisfies Fokker-Planck equation \cite{Risken_1989}
\begin{gather}\label{eq:kfp}
 \frac{\partial\mathcal G(P,t|P',0)}{\partial t}=\mathcal L(P,P^*)
\mathcal G(P,t|P',0),\\ \mathcal G(P,0|P',0)=\delta(P-P'),\nonumber\\\nonumber
\mathcal L(P,P^*)=\left[\i(\omega_0+\alpha|P|^2)+\frac{\Gamma_X}{2}\right]
\frac{\partial}{\partial P}P+c.c.+W\frac{\partial^2 }{\partial P\partial P^*}.
\end{gather}
Eq.~\eqref{eq:kfp} allows analytical solution for arbitrary pumping.\cite{Dykman1980} For relatively high pumping rate the spectrum 
is nonzero only for for $\omega-\omega_{\rm X}$ and
takes 
the  form
\begin{equation}\label{eq:spectrum_poisson}
 \av{|P(\omega)|^2}\propto W(\omega-\omega_{\rm X})\exp\left(-\frac{\omega-\omega_{\rm X}}{\alpha \av{n_{\rm X}}}\right), \omega>\omega_{\rm X}.
\end{equation}
We introduce the parameter $\av{n_{\rm X}}= W/\Gamma_{\rm X}$ which characterizes the stationary excitonic population $\av{|P(t)|^2}$, it depends both on pumping strength and on the excitonic decay rate, but it does not depend on the value of the non-linearity.
 It follows from Eq.~\eqref{eq:spectrum_poisson} that the emission spectrum of the system is strongly asymmetric, the maximum position is shifted by $\alpha\av{n_{\rm X}}$ from the non-interacting position. Moreover, the spectrum is strongly broadened: its width is of the same order as the energy shift. Eq.~\eqref{eq:spectrum_poisson} can be understood as follows. For the strong non-linearity, where $\alpha\av{n_{\rm X}} \gg \Gamma_{\rm X}$ the shape of the spectrum is determined by the fluctuations of the excitonic population $n_{\rm X}$.\cite{Dykman1980} The steady state distribution of $n_{\rm X}$ is the same as for $\alpha=0$, i.e. it is described by the Poisson formula.
In these case the root mean square value of fluctuations of $n_{\rm X}$ is
  proportional to the average population $\av{n_{\rm X}}$. 
 The spectrum is also asymmetric since interactions blue-shift exciton energy only, hence, high energy wing is larger than low energy one.

The very same considerations can be applied for the system of coupled dots and microcavity. The pumping results in the blue shifts of exciton energies and increase of the exciton resonance widths. As a result, the light-matter coupling becomes weaker. It is worth to stress, that the blue shift and broadening are contributed by all particles in the system, including the excitons generated in the higher energy states and in the wetting layer. For rather realistic pumping rates the blue shift can reach about $1$~meV (see, e.g., Ref.~\onlinecite{LeSiDang06}), hence the broadening of the exciton state due to the particle number fluctuations can exceed the Rabi slitting being about $0.5$~meV in the state-of-the-art structures.\cite{Peter2005} Therefore, the observation of the non-linear effects for quantum dots embedded into the microcavities may be hindered by the particle number fluctuations.



%

\end{document}